\documentclass[aps,superscriptaddress,prb,amsmath,amssymb,reprint,longbibliography]{revtex4-2}

\usepackage[USenglish]{babel}
\usepackage[T1]{fontenc}
\usepackage[utf8]{inputenc}
\usepackage{graphicx}
\usepackage{upgreek}
\usepackage[unicode]{hyperref}
\usepackage{multirow}

\begin{document}
%----------------------------------
	
\title{Shielding of external magnetic field by dynamic nuclear polarization \\in (In,Ga)As quantum dots}

\date{\today}
	
\author{E.~Evers}
\email{email: eiko.evers@tu-dortmund.de}
\affiliation{Experimentelle Physik 2, Technische Universit\"at Dortmund, 44221 Dortmund, Germany}
	
\author{N.~E.~Kopteva}
\affiliation{Experimentelle Physik 2, Technische Universit\"at Dortmund, 44221 Dortmund, Germany}
\affiliation{Spin Optics Laboratory of St.\,Petersburg State University, 198504 St.\,Petersburg, Russia}

\author{I.~A.~Yugova}
\affiliation{Spin Optics Laboratory of St.\,Petersburg State University, 198504 St.\,Petersburg, Russia}
\affiliation{Physical Faculty of St.\,Petersburg State University, 198504 St.\,Petersburg, Russia}

\author{D.~R.~Yakovlev}
\affiliation{Experimentelle Physik 2, Technische Universit\"at Dortmund, 44221 Dortmund, Germany}
\affiliation{Ioffe Institute, Russian Academy of Sciences, 194021 St.\,Petersburg, Russia}

\author{M.~Bayer}
\affiliation{Experimentelle Physik 2, Technische Universit\"at Dortmund, 44221 Dortmund, Germany}
\affiliation{Ioffe Institute, Russian Academy of Sciences, 194021 St.\,Petersburg, Russia}

\author{A.~Greilich}
\affiliation{Experimentelle Physik 2, Technische Universit\"at Dortmund, 44221 Dortmund, Germany}

\begin{abstract}
The dynamics of the coupled electron-nuclear spin system is studied in an ensemble of singly-charged (In,Ga)As/GaAs quantum dots~(QDs) using periodic optical excitation at 1\,GHz repetition rate. In combination with the electron-nuclei interaction, the highly repetitive excitation allows us to lock the electron spins into magnetic resonance in a transverse external magnetic field. Sweeping the field to higher values, the locking leads to an effective ‘‘diamagnetic'' response of significant strength due to dynamic nuclear polarization, which shields the QD electrons at least partly from the external field and can even keep the internal magnetic field constant up to 1.3\,T field variation.
We model the effect through a magnetic field-dependent polarization rate of the nuclei, from which we suggest a strategy for adjusting the nuclear polarization through the detuning between optical excitation and electronic transition, in addition to tuning the magnetic field.
\end{abstract}

\maketitle

\section{Introduction}

The spin degree of freedom of charge carriers localized in semiconductor quantum dots (QDs) continues to attract considerable interest due to the large optical dipole moments and fast optical spin control. These properties allowed, for example, entanglement between distant spins~\cite{StockillPRL17}, relevant for quantum repeater protocols, or generation of cluster states of entangled photons, a possible resource for quantum computation~\cite{Schwartz434}. However, to be competitive with other platforms~\cite{Ladd2010}, the spin coherence has to be extended. The main source of QD carrier spin decoherence is the coupling to the surrounding nuclear spins~\cite{MerkulovEfrosRosen,Dyakonov,Urbaszek2013,Glazov_book}. A way to improve the coherence properties is polarization of the nuclear spins. However, to have a substantial impact, a very high degree of polarization close to 100\% is needed~\cite{LossPRB04}. Such a nuclear polarization can also be used for a long-living quantum memory to store the coherent state of an electron spin~\cite{TaylorPRL03,SchwagerPRB10}. A coherent coupling between an electron spin and a collective nuclear spin excitation (spin wave) was recently demonstrated~\cite{GangloffScience19,GangloffArxiv2020}. Further theoretical investigation suggested that about 50\% of nuclear polarization is sufficient to reach up to 90\% fidelity in quantum state storage with strain-enabled nuclear spin waves~\cite{LeGallPRL19}.

Rather high values of nuclear spin polarization were reported in literature~\cite{BrackerPRL05,EblePRB06,ChekhovichPRL10,Chekhovich2017}, reaching up to 65\% in InAs/GaAs QDs~\cite{ChekhovichPRL10} and up to 80\% in strain-reduced GaAs/(Al,Ga)As QDs~\cite{Chekhovich2017}. It is important to note that all these studies were performed on single QDs subject to a longitudinal magnetic field~\cite{Urbaszek2013}.

In the present paper, we demonstrate the experimental implementation of a protocol for achieving high nuclear polarization in an inhomogeneous ensemble of QDs subject to a transverse magnetic field. Advancing the methods demonstrated on single QDs~\cite{Latta2009,VinkNatPhys09} and analyzed in Refs.~\cite{Carter,Korenev2011}, we exploit electron spin magnetic resonance due to locking by nuclear polarization, based on highly repetitive pulsed laser excitation that reduces the nuclear spin fluctuations. After locking the whole QD ensemble to single frequency precession about the magnetic field~\cite{Evers2021}, we sweep the external magnetic field and drag the precession away from the precession frequency corresponding to the momentary field. During this sweep a high nuclear polarization develops, leading to a strong diamagnetic response by shielding partly the Zeeman interaction with the external magnetic field for the confined electrons. Simultaneously, we pick up all magnetic resonance modes in the precession spectrum between the start and end fields.

\section{Sample and setup}
\begin{figure*}[t]
	\begin{center}
		\includegraphics[trim=0mm 1mm 1mm 1mm, clip, width=\textwidth]{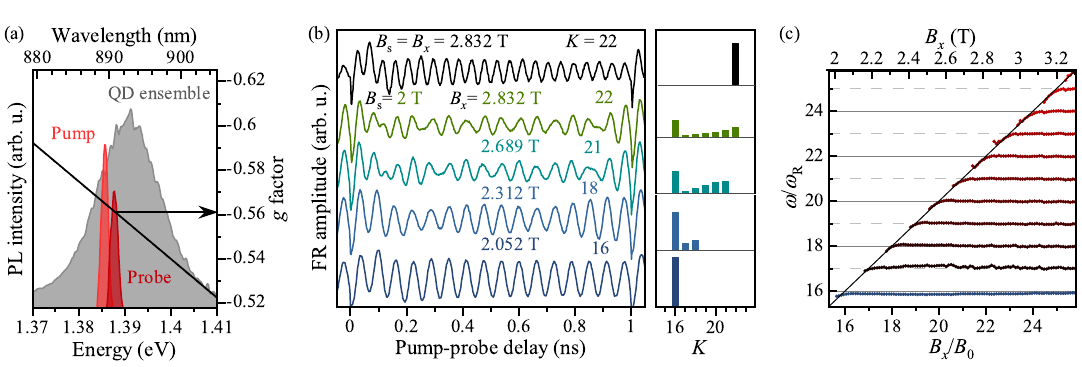}
		\caption{(a) Spectra of the QD ensemble PL (gray) and laser emission of pump and probe (red). Black line gives the $g$-factor dispersion. The central emission energy for the pump is at $1.3867\,$eV and for the probe at $1.3878\,$eV for the laser with $1\,$GHz repetition frequency. The $75.76\,$MHz repetition frequency laser in Sec.~\ref{sec:two-laser} is degenerate for pump and probe at $1.3867\,$eV central photon energy. (b) Faraday rotation signals for several $B_{x}$. At $2.832\,$T, nominally corresponding to $K=22$ (see text), one observes six beating nodes between pulse arrivals in the time trace, demonstrating that the signal is contributed by seven frequencies with different $K$. The contributing mode amplitudes vs $K$ are shown in the right panel for the different $B_{x}$. (c) Frequencies of the cosine components (marked by differing colors) with a larger than 1$\,\%$ amplitude relative to the FR signal amplitude against the external magnetic field $B_{x}$ in absolute units (top axis) and in units of mode number $B_{x}/B_0$ (bottom axis) with the magnetic field sweep started from $B_{\rm s}=2$\,T. The different frequencies can be given starting from the magnetic field in the sweep from which on they contribute. The black diagonal line shows the field dependence of the precession frequency without nuclear contribution. $E_{\text{pump}}=1.3867\,$eV, $E_{\text{probe}}=1.3878\,$eV, and $\Delta=-0.53$.}
		\label{fig1}
	\end{center}
\end{figure*}

The sample was grown by molecular beam epitaxy on a (100)-oriented GaAs substrate and consists of 20 layers of $10^{10}$\,cm$^{-2}$ (In,Ga)As QDs separated by $80\,$nm GaAs layers~\cite{Greilich2006}. A silicon $\delta$-doping layer was placed $16\,$nm above each QD layer, providing on average a single resident electron per dot. The sample was thermally annealed at $945\,^{\circ}$C to homogenize the QD size distribution and to shift the central QD emission energy to 1.39\,eV. The indium to gallium ratio in the QDs is approximately 35 to 65~\cite{Petrov2008}. The sample is placed in a cryostat at the temperature of $5.3\,$K and an external magnetic field ($B_{x}$) applied transverse to the light propagation (Voigt geometry). The photoluminescence (PL) spectrum (gray shaded area) together with the $g$-factor dispersion (solid black line) can be found in Fig.~\ref{fig1}(a).

The key element in the setup is a Ti:Sapphire laser emitting pulses of $150\,$fs pulse duration with $1\,$GHz repetition frequency, equivalent to a pulse separation $T_{\text{R}} = 1\,$ns. The pump and probe beams pass through individual spectral filters, formed by a grating and a slit. Thereby we reduce the pulse spectral width to about $2\,$meV and extend the duration to 2\,ps, see the red shaded areas in Fig.~\ref{fig1}(a). A mechanical delay line provides the variable time delay $t$ between pump and probe pulses. A double modulation scheme is implemented by modulating the pump beam between left- and right-circular polarization at $84\,$kHz, while the probe beam is intensity modulated at $100\,$kHz for fixed vertical polarization. The pump beam is focused to a diameter of $50\,\mu$m, the probe beam to 40\,$\mu$m. The average laser power is $4\,$mW for the pump and $0.15\,$mW for the probe beam. An optical polarization bridge is used to measure the Faraday rotation~(FR) of the transmitted probe behind the sample.

\section{Experimental results}
\subsection{Mode dragging and mode pickup}
We start the discussion with experimental evidence of two phenomena:  \textit{mode dragging} and \textit{mode pickup}. The top black curve in Fig.~\ref{fig1}(b) demonstrates the pump-probe trace at the external field $B_{x}=2.832\,$T, where the sample was kept in darkness during ramping up to that field strength. The signal shows periodic oscillations with a single Larmor frequency corresponding to the electron $g$ factor of $g_{\rm e}=-0.56$. The pulsed excitation with 1\,GHz rate induces a magnetic resonance by forcing the Larmor precession of all optically addressed resident electron spins to occur on a single mode due to nuclear induced frequency focusing~\cite{Evers2021}. The frequency $\omega$ of this mode is commensurate to the laser repetition frequency $\omega_{\text{R}}=2\pi/T_{\rm R}$: $\omega_K=K\omega_{\text{R}}$, with $K$ being the integer that gives the number of full spin revolutions about the magnetic field between two laser pulses. We use $K$ for characterizing the commensurate precession modes. The magnetic fields, where the commensurability condition is fulfilled can be expressed also as $B_{x} = K B_0$, as $\omega = g_{\text{e}} \mu_{\text{B}} B_{x}/\hbar$. For our sample, this results in $B_0 = 2\pi \hbar / (T_{\rm R} g_{\text{e}} \mu_{\text{B}}) = 128$\,mT with the Bohr magneton $\mu_{\text{B}}$ and the reduced Planck constant $\hbar$. 

In contrast, the green trace in Fig. \ref{fig1}(b) demonstrates the measurement at the same field of $B_{x}=2.832\,$T, but starting from $B_{\rm s}=2\,$T during the magnetic field sweep the sample was illuminated with the pulsed laser. The field ramping speed was $6.17\,$mT/s. Surprisingly, the pump-probe trace is no longer the expected single harmonic but shows a pronounced beating pattern with six beating nodes between two pump pulses. The beating has to arise from the superposition of several discrete precession modes, where for six nodes seven different frequencies have to contribute. To shed more light on this result, we perform measurements where the field sweep was stopped at lower magnetic fields, corresponding to particular $K$, while the illumination also started at 2\,T. At 2.052\,T, corresponding to $K=16$, the FR signal represents again a single harmonic. Increasing the external field from there in units of about $B_0$, so that the nominal $K$ is increased in steps of unity, the time-traces contain a proportionally increasing number of beating nodes.

For a more detailed analysis, we fit each signal with a superposition of cosine functions:
\begin{equation}
S(t) = \sum_i S_{0,i} \cos(\omega_i t),
\label{eq:S}
\end{equation}
where $S_{0,i}$ is the amplitude of the mode with oscillation frequency $\omega_i$. Generally, the contained frequencies indeed correspond to those of commensurate modes. The fitted frequencies as a function of $K$ for the different measurement magnetic fields are shown in Fig.~\ref{fig1}(c). The 1\,GHz~excitation leads to single frequency focusing onto the closest mode at a fixed $B_{x}$, so that one would expect that the precession frequency jumps continuously from one single mode to the next higher lying one when the field is swept, leading to a steplike behavior~\cite{Evers2021}. However, unexpectedly the initial single frequency mode component ($K=16$) is present over the whole swept field range. We call this effect {\it mode dragging}, as this mode is maintained even though the nominal precession frequency corresponding to $B_{x}$ steadily increases. Additionally, by sweeping through the higher modes with $K>16$ they become also activated and dragged. We term this effect  {\it mode pickup}. Starting from the $K=16$ mode, the precession spectrum collects all the commensurate modes that are located in the range between the start and end field of the sweep. Whenever the magnetic field reaches a value allowing for precession with one more revolution, this mode appears in the spectrum. In Sec.~\ref{sec:two-laser}, we present two-laser pump-probe traces, which allow us to resolve all precession modes via a direct fast Fourier transformation and support our fitting approach with multiple discrete frequencies.

\begin{figure}[t]
\begin{center}
		\includegraphics[width=\columnwidth, trim=3mm 4mm 3mm 4mm]{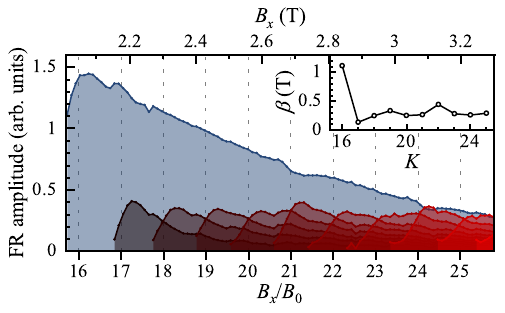}
		\caption{Mode amplitudes in the precession spectrum ($S_{0,i}$) vs $B_{x}$, obtained from a fit with a superposition of cosine functions [Eq.~(\ref{eq:S})]. The inset shows the diminishing constant ($\beta$) for each mode. $E_{\text{pump}}=1.3867\,$eV, $E_{\text{probe}}=1.3878\,$eV, and $\Delta=-0.53$.}
		\label{fig2}
	\end{center}
\end{figure}

Obviously, an electron in a particular quantum dot precesses on a single mode. The extended mode spectrum is an ensemble effect, where each mode amplitude represents the fraction of dots in the ensemble with the associated precession frequency. That this frequency stays for large dot fractions below that corresponding to the end field of the sweep, means that this field is partly shielded. The largest dot fraction remains even stuck at the precession frequency of the sweep start and does not experience the increase of $B_{x}$ by 1.3\,T at all; the field strength increase from $2\,$T up to $3.3\,$T is totally compensated.

The evolution of the mode amplitudes $S_{0,i}$, defined by Eq.~(\ref{eq:S}), with increasing magnetic field is shown partly as a bar diagram in the Fig.~\ref{fig1}(b) and in fine steps in Fig.~\ref{fig2}. Across the whole sweep, the $K=16$ mode as the first to fulfill commensurability at $2.052\,$T (shown by the blue shaded area) has the highest amplitude and decays slowly with increasing magnetic field. Whenever an additional commensurate mode frequency is crossed in the field sweep, its amplitude rises fast with $B_{x}$ across a field range comparable to the nuclear spin fluctuations and reaches its maximum value close to $B_{x} = K B_0$ before dropping with a further increase of the field. One can extract a characteristic diminishing field ($\beta$) for each mode, using an exponential fit to the $S_{0,i}$ dependence on $B_{x}$. The inset in Fig.~\ref{fig2} shows $\beta$ for the different modes $K$. The largest diminishing field belongs to the lowest mode ($K=16$) with $\beta = 1.12\,$T, all other modes diminish across a shorter range of $\beta = 0.3$\,T.

Using the factor $\beta=1.12$\,T, one can estimate the fraction of QDs, that would reach a 100\% nuclear polarization, once being locked into a single mode at a start magnetic field, see Appendix~\ref{app1} and Eq.~(\ref{eq:max-field}). This fraction would be about $0.4\,$\%, or 400\,QDs, assuming $10^5$ initially locked QDs. Experimentally, we demonstrate values of the Overhauser field up to 1.5\,T (that corresponds to about $24\,$\% nuclear polarization), as shown in Sec.~\ref{sec:two-laser}, which, however, does not represent the ultimate limit, as parameters like the magnetic field sweep range as well as the excitation parameters can be further optimized.

\subsection{Mode dragging for sweeping towards lower magnetic fields}\label{field-down}
\begin{figure*}[t]
\begin{center}
		\includegraphics[width=\textwidth, trim=2mm 2mm 2mm 2mm]{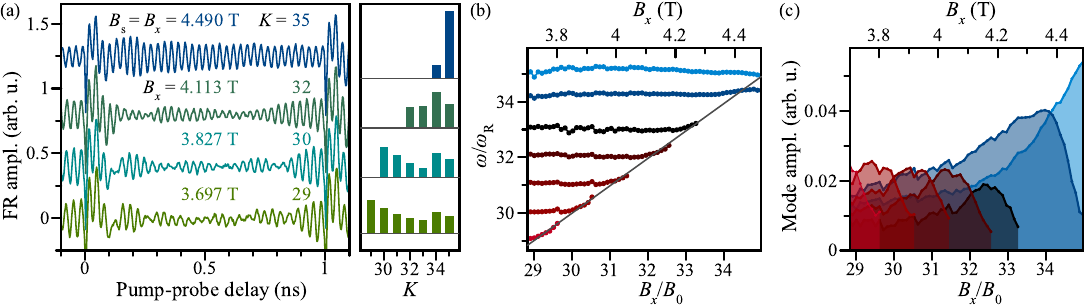}
		\caption{Spin mode dragging towards lower magnetic fields. (a) Pump-probe spectra at different external fields $B_x$, resonant to different modes $K=B_x/B_0$, starting from $B_{\rm s}=4.49 \,$T, which corresponds to $K=35$. For lower fields beating nodes show up in the Faraday rotation patterns, evidencing the contribution of discrete Larmor frequencies. These Larmor frequencies with their respective amplitudes are shown in the right panel. (b) Frequencies of the constituent cosine components with a relative amplitude larger than $1\,\%$ of the total FR amplitude as function of the external magnetic field $B_{x}$ in absolute units (top axis) and in units of mode number $K=B_{x}/B_0$ (bottom axis). Only discrete Larmor frequency modes $\omega/ \omega_{\text{R}}$ are found and show dragging when decreasing the magnetic field for persistent illumination with the 1\,GHz laser. Black diagonal line shows the field dependence of the precession frequency without nuclear contribution. (c) Amplitude dependence of the contributing precession modes on the external field. $E_{\text{pump}}=1.3867\,$eV, $E_{\text{probe}}=1.3878\,$eV, and $\Delta=-0.53$.}
		\label{fig3}
	\end{center}
\end{figure*}
Multiple modes also appear in the precession spectrum when the external magnetic field is decreased under continuous illumination with the $1\,$GHz laser. By dragging electron spins similarly as described above, the precession modes activated during the sweep at higher magnetic fields are also present in lower magnetic fields, see Fig.~\ref{fig3}(a). The left panel shows pump-probe traces for different external fields $B_{x}$ measured one after the other in a series of measurements started from the field $B_{\rm s}=4.49\,$T. At the starting field the preferred mode is that with number $K=35$ and only this mode is present. When sweeping the field down to $B_{x}=4.113\,$T with a speed of $6.17\,$mT/s under 1\,GHz illumination, we observe the beating structure with nodes, as shown for the external fields nominally corresponding to $K=32\text{,}\,30\text{,}\,29$. The number of contributing precession modes corresponds to the number of modes swept over by the external field. The bars in the right panel represent these modes. Figure~\ref{fig3}(b) shows the frequencies contained in the pump-probe traces, determined by a superposition of harmonic functions with laser-commensurate frequencies as in the case of the increasing field. These modes are persistent over the whole scanned external field range. The amplitudes of the modes are shown in Fig.~\ref{fig3}(c). The picture is similar to the case with increasing external field, except of the inversion with respect to the stating mode corresponding to the sweep start field $K=35$. This mode is basically the only one at $B_{\rm s}=4.49\,$T as it is the nuclei-focused mode at this field; its amplitude decreases slowly during the sweep with a diminishing constant $\beta=0.5\,$T. The mode number 34 is already present and builds up strongly before it reaches its maximum at the field of $B_{x}=4.363\,$T, then it decreases with $\beta=0.6\,$T. The other modes decrease with the same $\beta$, similar to the case of increasing external field. We attribute the small differences in $\beta$ to the difference in starting field being not exactly on a preferred field so that neither mode 35 nor mode 34 can reach their maximum polarization and hence nuclear fluctuation narrowing. While the case for increasing external field leads to an Overhauser field antiparallel to the external field, decreasing the external field establishes an Overhauser field parallel to the external field. The bidirectionality of the Overhauser field is consistent with the findings for the previously described spectral dragging~\cite{Hoegele2012} and anomalous Hanle effect~\cite{Krebs2010}.

\subsection{Direct measurement of the mode structure}\label{sec:two-laser}
\begin{figure}[t]
	\begin{center}
		\includegraphics[trim=0mm 1mm 1mm 1mm, clip, width=\columnwidth]{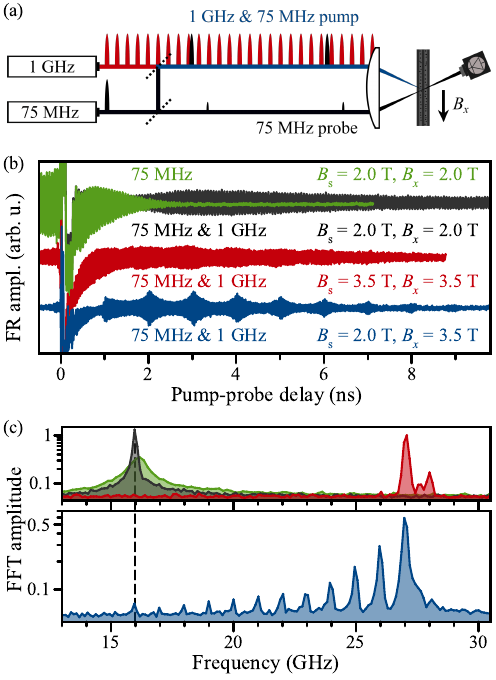}
		\caption{(a) Experimental scheme of the two laser setup. The $1\,$GHz (red) and the $75.76\,$MHz (black) pump pulses have an arbitrary delay towards each other, changing for every pulse as their repetition frequency is not commensurable. The probe pulses maintain a constant delay to the $75.76\,$MHz pump pulses. The sample is illuminated in growth direction and placed in an external field $B_x$ perpendicular to it. (b) Pump-probe FR for pulsed excitation with $75.76$\,MHz only (green) and together with the 1\,GHz laser at $B_{x} = 2\,$T (black) and $B_{x} = 3.5\,$T (red). Blue trace shows the signal after a magnetic field sweep from $B_{\text{s}}=2\,$T to $B_{x}=3.5\,$T at a rate of $6.17\,$mT/s. During sweeping, only the $1\,$GHz laser was on. (c) FFT spectra of the corresponding traces, note the logarithmic scale. The blue FFT spectrum shows 12 precession modes with the lowest one corresponding to the precession frequency at the field of $2\,$T, while measuring at an external field of $3.5\,$T.}
		\label{fig4}
	\end{center}
\end{figure}
Possible doubts could come along with fitting a complex oscillatory pattern with a superposition of a few cosine functions having frequencies commensurable to the $1\,$GHz repetition rate to a signal of only $1\,$ns duration. To give further experimental evidence of the discrete precession modes we resolve them by increasing the measurement time using a $75.76\,$MHz pulsed laser only, as well as in combination with the $1\,$GHz laser applied simultaneously for excitation [see Fig.~\ref{fig4}(b)]. The additional laser emits pulses with a pulse separation of $13.2\,$ns and has a spectral width of $2\,$meV full width at half maximum (FWHM) with a pulse duration of $2\,$ps. We can introduce an optical detuning between the $75.76\,$MHz laser and the $1\,$GHz laser, see the red colored peaks in Fig.~\ref{fig1}(a) for the pulses of $2\,$ps duration. Only the pump pulses of the $1\,$GHz laser are applied at the pump energy, while the pump and probe pulses of the $75.76\,$MHz laser are used at the probe frequency with degenerate pump and probe energies. A mechanical delay line introduces the delay between the pump and probe pulses of the $75.76\,$MHz laser. The lasers are not synchronized or phase-locked to each other. Thus, only electron spins oriented by the $75.76\,$MHz pump pulses are measured by the $75.76\,$MHz probe pulses and the influence of the electron spins oriented by the $1\,$GHz pump pulses is only visible via their impact on the nuclear spins. We use the already described double modulation scheme and direct the $1\,$GHz laser excitation beam through the pump path as depicted in Fig.~\ref{fig4}(a). The typical average beam powers are $4\,$mW and $0.15\,$mW for the low repetition laser pump and probe and $7\,$mW for the high repetition laser.

The green trace in Fig.~\ref{fig4}(b) demonstrates the pump-probe measurement using the $75.76\,$MHz laser at $B_{x} =2\,$T. The signal shows periodic oscillations with an average Larmor frequency corresponding to the resident electron $g$ factor $g_{\text{e}}=-0.56$ [see Fig.~\ref{fig1}(a)], which shows a dephasing time of $T_2^*=1.4\,$ns~\cite{Greilich2006a}. By adding the additional pump with $1\,$GHz laser repetition frequency to the same QD ensemble, the dephasing is elongated significantly which is related to the reduction of the precession frequency spread of the excited QD ensemble down to the single-mode regime and a reduction of the nuclear spin fluctuations~\cite{Evers2021}, see the black trace in Fig.~\ref{fig4}(b). Note that the signal originates from the electrons excited by the $75.76\,$MHz laser only. The modulation in the black trace is coming from the overlap of the fast decaying signal of QDs excited only at 75.76\,MHz without reduced frequency spread and those with reduced frequency spread. The fast Fourier transform (FFT) of the corresponding time traces shows single peaks, see the green and black peaks in Fig.~\ref{fig4}(c). The broad-peak amplitude of the black spectrum is reduced in comparison to the green one, and an additional narrow peak develops at the frequency of 16\,GHz. The red traces in Figs.~\ref{fig4}(b) and~\ref{fig4}(c) correspond to $B_{x}=3.5$\,T. For all spectra the laser excitation was blocked until the final magnetic field was reached.

In contrast, the blue trace in Fig.~\ref{fig4}(b) demonstrates the situation when the magnetic field was first increased from 2\,T to $3.5\,$T at a rate of $6.17\,$mT/s, with simultaneous continuous excitation using the $1\,$GHz laser. The FR measured by the $75.76\,$MHz pump-probe laser at 3.5\,T shows now beatings in the time trace, with signal outbursts that are separated by $1\,$ns. The FFT of the signal shows 12 peaks, each separated by $1\,$GHz, see the blue spectrum in Fig.~\ref{fig4}(c). As one can see, the lowest frequency component of the signal corresponds to the precession frequency at the field of $2\,$T, while the applied external field is $3.5\,$T.

\subsection{Additional findings}
To bring down a round figure, we summarize additional observations: (a) The sum of all mode amplitudes in the precession spectrum stays constant across the sweep, as expected from the number of QDs optically excited, see Fig.~\ref{fig1}(b). The number of the polarized electron spins stays constant and the precession frequencies redistribute in the electron spin ensemble with varied external field. (b) After the end of the sweep, the low frequency modes persist under pulsed illumination for at least one hour, and decay on a timescale of seconds to minutes in darkness (see Appendix~\ref{app:times}). As the focusing of the precession modes on commensurate frequencies arises from the nuclei generating the Overhauser field that acts on the confined QD electron, we attribute the remanence effects to the nuclei as well, supported also by the involved time scales. The maximal nuclear polarization for the studied QD ensemble can in principle go up to $B_{\text{N,max}}=6.23$\,T (see Appendix~\ref{app1}).\\

\section{Model results}
To describe the observed effects, we represent the coupled electron-nuclear spin dynamics due to pulsed periodic optical excitation in the frame of a dynamical nuclear polarization model~\cite{Korenev2011}. 

\subsection{Overhauser field under pulsed optical excitation}
\begin{figure*}[t]
	\begin{center}
		\includegraphics[trim=0mm 0mm 0mm 0mm, clip, width=2.05\columnwidth]{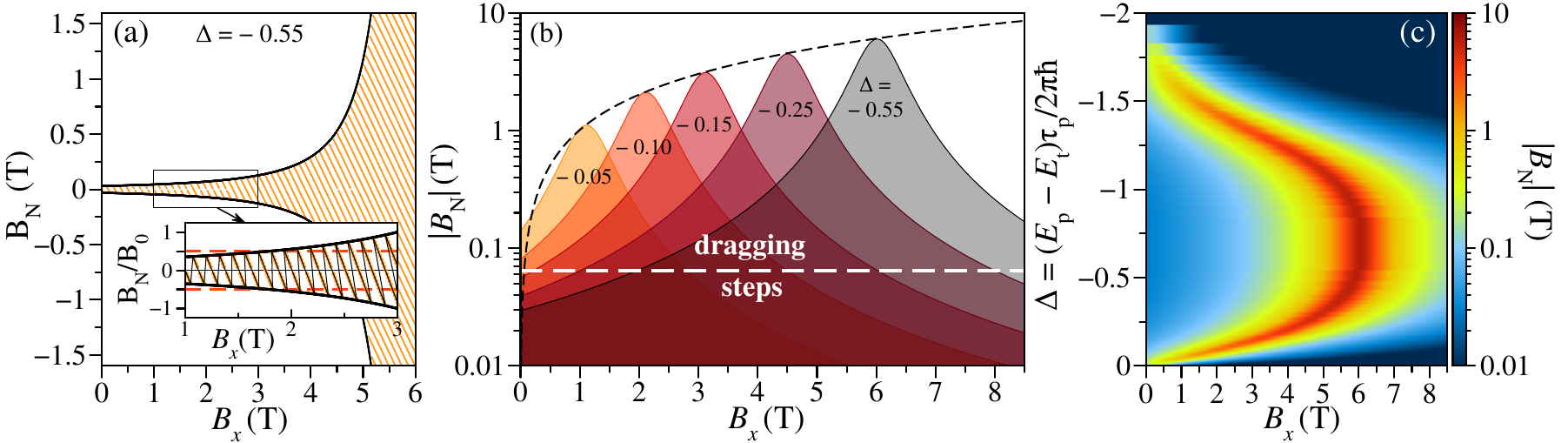}
		\caption{(a) Variation of the Overhauser field with the variation of $B_x$, shown by the orange dots. The black lines are the envelopes, connecting the maximal values of the $B_{\rm N}$. Inset shows the internal structure in the range of fields 1 to 3\,T. Red dashed lines show the border between the steplike and dragging regimes. (b) Dependence of the Overhauser field $B_{\rm N}$ on $B_{x}$ for several values of detuning $\Delta$. White dashed line gives the border $B_{\rm N} = B_0/2$ between steplike behavior and mode dragging and pickup. (c) Contour plot of $B_{\rm N}$ vs $\Delta$ and $B_{x}$ in logarithmic scale. Used parameters: $T_{\rm R} = 1$\,ns,  $T_{2,0} = 2$\,$\mu$s, $\tau_{\text{c}} = 10$\,ns, $\theta = \pi$, $\Delta = - 0.55$, $A = 49.2$ $\mu$eV, $\bar{Q} = 20$, $N = 5.5 \times 10^5$, $T_\text{1L} = 2$\,min, $f_\text{N0} = 0.86$, and $g_{\rm e}=-0.57$.}
		\label{fig5}
	\end{center}
\end{figure*}
The electron spin component ($S_x$) acting against the external magnetic field for an up-field sweep is generated by the optically induced Stark shift due to a nonzero optical detuning of the pump pulses from the probed trion resonance. In case of excitation by an infinitely long sequence of circularly polarized pulses, the averaged (over the laser repetition time)  $S_x$ component of the electronic polarization is given by~\cite{Yugova2009}:
\begin{eqnarray}
\label{eq:Sx}
\langle S_{x}(t)\rangle&=&\frac{T_\text{2}}{T_\text{R}} \left[1-\exp\left(-T_\text{R}/T_\text{2}\right)\right] \cdot \\ \nonumber &\cdot&\frac{K(1-Q^2)\sin(\omega T_\text{R})}{4(1+LM - (L+M)\cos(\omega T_\text{R}))}.
\end{eqnarray}
Here
\begin{gather}
K = \frac{Q\exp{(-T_\text{R}/T_\text 2)}\sin \Phi}{1 - Q\exp{(-T_\text{R}/T_\text{2})}\cos \Phi},\\
M = Q(\cos \Phi - K \sin \Phi) \exp{(-T_\text{R}/T_\text{2})},\\
L = \frac{1 + Q^2}{2}\exp{(-T_\text{R}/T_\text{2})}.
\end{gather}
$T_\text{2}$ is the homogeneous electron spin relaxation time in a transverse external magnetic field, $\omega_{\text{e}} = \mu_\text{B} g_{\rm e} B_{x}/\hbar$ is the electron Larmor precession frequency in the external magnetic field, and $\omega_{\text{N}} = \mu_\text{B} g_{\rm e} B_\text{N}/\hbar$ is the electron spin precession frequency in the created Overhauser field, which changes the precession frequency of the electron to $\omega = \omega_{\text{e}}+\omega_{\text{N}}$. $Q$ and $\Phi$ are the characteristics of the optical pulse, depending on the dimensionless optical detuning $\Delta=(E_\text{p}-E_\text{t})\tau_p/(2\pi\hbar)$ between the laser photon energy $E_\text{p}$ of the pump and the trion transition energy $E_\text{t}$, with the pump-laser-pulse duration $\tau_p$ and $\theta$ is the optical pulse area~\cite{Yugova2009}.

The dynamic nuclear spin polarization directed along or opposite to the external magnetic field can be described by a transcendental equation where the nuclear spin polarization depends on the electron spin polarization and vice versa~\cite{Abragam1961,Korenev2011}:
\begin{equation}\label{eq:DNP}
\frac{dI_\text{N}}{dt} = - \frac{1}{T_\text{1e}} [I_\text{N} - \bar{Q} \langle S_{x}(I_\text{N})\rangle] - \frac {I_\text{N}}{T_\text{1L}}.
\end{equation}
$I_\text{N}$ is the average nuclear spin polarization and $\bar{Q} = \sum_{j}4I_j(I_j+1)n_{\text{QD},j}/3$ is a factor that depends on the nuclear spin averaged over all nuclei species with spin ($I_j$) and the fraction $n_{\text{QD},j}$ for the elementary cell with two atoms. $\bar{Q} = 20$ for (In,Ga)As/GaAs QDs with 35\,\% of In concentration. $T_\text{1e}$ is the nuclear spin polarization time due to electron polarization and $T_\text{1L}$ is the nuclear spin-lattice relaxation time, which takes into account any other possible leakage like spin diffusion. The proportion of these times can conveniently be summarized in the leakage factor $f_\text{N} = T_\text{1L}/(T_\text{1L} + T_\text{1e})$. The nuclear polarization produces a nonzero Overhauser field $B_{\rm N} = AI_\text{N}/\mu_{\rm B} g_{\rm e}$ which acts back on the electron spins. The hyperfine constant $A = 49.2$\,$\mu$eV is averaged over all nuclear species with respect to their natural abundance and fraction in the QDs. This leads to a nuclear polarization, which compensates the external magnetic field change and locks the electron Larmor frequency $\omega$ to an integer mode of the laser repetition frequency $\omega_{\text{R}}$. This model was confirmed by several experiments~\cite{Carter,Zhukov2018,MarkmannNatCom19,Evers2021}.

The nuclear polarization rate depends on the external magnetic field and the average nuclear spin polarization and can be written as~\cite{EblePRB06,Glazov_book}:
\begin{equation}\label{eq:T1N}
\frac{1}{T_\text{1e}} = \left(\frac{A}{\hbar N}\right)^2 \frac{2F\tau_{\text{c}}}{1+(\omega_{\text{e}} + \omega_{\text{N}})^2\tau_{\text{c}}^2}.
\end{equation}
$N$ is the number of nuclei, $\tau_{\text{c}}$ is the correlation time in the electron-nuclear spin system, the factor $F$ represents the average fraction of time during which the dot is occupied. As can be seen from Eq.~(\ref{eq:T1N}), the shortest nuclear spin polarization time corresponds to the case when the Overhauser field compensates for the external magnetic field, or $\omega_{\text{e}} = -\omega_{\text{N}}$. It shows that the spin flip-flop process in the electron-nuclear spin system is most efficient when the Zeeman splitting of electron spin levels is comparable with the Zeeman splitting of the nuclear spin levels. Due to the large difference in Bohr magnetons for nuclei and electrons in an external magnetic field this process is suppressed. However, its efficiency can be increased if the Zeeman splitting of electron levels is compensated by the Overhauser field. The efficiency is maximal when the Overhauser field reaches the value of the external magnetic field.

As the initial step we fix the experimental parameters and calculate the value of $\langle S_x\rangle$. For the fixed value of $\langle S_x\rangle$ we calculate the average nuclear spin polarization $I_\text{N}$. From this we further calculate the electron precession frequency in the Overhauser field $\omega_\text{N}$. This value is used in Eq.~(\ref{eq:DNP}) to calculate the dependence of the nuclear spin relaxation time $T_\text{1e}$, and, as a result, of the leakage factor $f_\text{N}$, as a function of the external magnetic field (or $\omega_\text{e}$). With the calculated dependence of the $f_\text{N}(\omega_\text{e}+\omega_\text{N})$ we find the stationary solutions of Eq.~(\ref{eq:DNP}) for the pulsed excitation. $f_\text{N}$ and $\langle S_x\rangle$ depend on $I_\text{N}$ and vice versa. As a result we get a set of discrete values of electron spin precession frequencies $\omega=\omega_\text{e} + \omega_\text{N}$ for different external magnetic fields.

Figure~\ref{fig5}(a) demonstrates the evolution of the Overhauser field $B_{\rm N}$ as a function of the rising external field $B_x$. The orange dots are the calculated values, while the thick black lines connect the maximal Overhauser field values (the so-called envelope). The inset of Fig.~\ref{fig5}(a) shows a zoom on a range of $B_x$ from 1 to 3\,T and allows one to see the details of the saw-tooth like behavior of the Overhauser field amplitude which builds up for a compensation of the changing values of $B_x$. For the range of $B_x<2$\,T and for the chosen optical detuning of $\Delta = -0.55$, the Overhauser field has its maximum values below $B_0/2$, due to the small transfer rate of electron polarization to the nuclei. It means, that for $B_x<2\,$T the locking of the Larmor frequency is only possible within the mode separation ($B_0/2=64$\,mT), see the red dashed lines at $|B_\text{N}| = |B_0|/2$.

At $B_x > 2$\,T, due to a higher polarization transfer rate, the values of $B_{\rm N}$ can increase above $B_0/2$ and neighboring frequency modes start to overlap. Therefore, one can observe a set of stable discrete solutions corresponding to different Overhauser fields, which lead to the presence of multiple frequencies at which the system can stay for a given $B_x$. The envelope connecting the maximal values of Overhauser fields [shown by black line in Fig.~\ref{fig5}(a)] is determined by the variation of the leakage factor with the external magnetic field and changes nonlinearly with it. Furthermore, the value of the Overhauser field depends on the optical detuning $\Delta$, as it is shown by the envelope functions in Fig.~\ref{fig5}(b) for different values of $\Delta$. For $\Delta = -0.55$ in Fig.~\ref{fig5}(b) the amplitude of $B_{\rm N}$ created by the compensation of $B_{x}$ can reach up to 6\,T. Additionally, we can conclude, that the maximal $B_{\rm N}$ depends on the detuning $\Delta$, as seen from the $B_\text{N}(\Delta,B_x)$ dependence, shown by the black dashed line running through the maxima of all curves. We summarize these simulations in the contour plot of Fig.~\ref{fig5}(c).

In the experiments, we denote the starting field of the sweep as $B_{\rm s}$, and the ramping to $B_{\rm s}$ is done without illumination of the QDs, which enables a build-up of the Overhauser field $B_{\rm N}$ starting from $B_{\rm s}$ only. If $B_{\rm s}$ would be set to 5\,T, the maximal $B_{\rm N}$ could reach only $\sim$1\,T by sweeping up the field. If $B_{\rm s}$ would be set to 0\,T, the maximal Overhauser field would also not reach the full 6\,T by sweeping the external field above $B_x=6\,T$. This is due to the fact, that up to $B_x<2\,$T the maximal Overhauser field cannot overcome the value of $B_0/2$ and only starting from 2\,T it becomes possible to polarize nuclei above $B_0/2$.

\begin{figure}[t]
	\begin{center}
		\includegraphics[trim=0mm 1mm 0mm 1mm, clip, width=1.05\columnwidth]{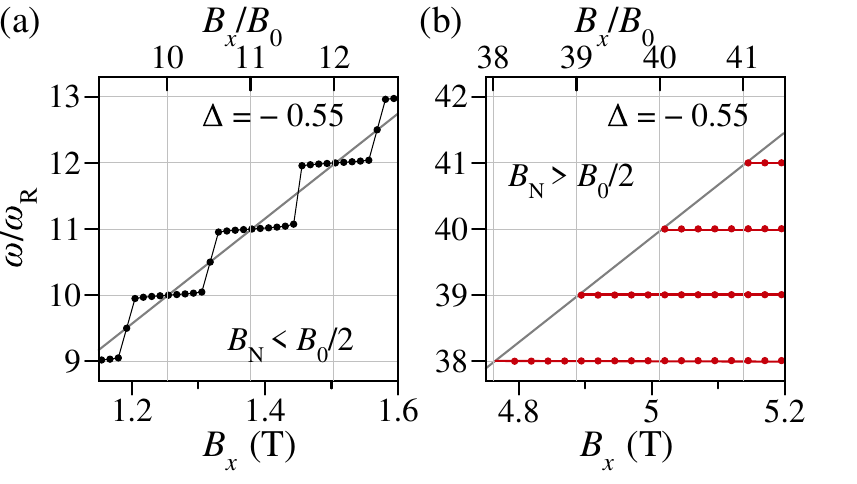}
		\caption{Ensemble precession frequencies vs transverse magnetic field swept starting from $B_{\rm s}=1.1$\,T~(a) and $B_{\rm s}=4.77$\,T~(b). Gray diagonal lines give $\omega_{\text{e}}$ without nuclear effects. Used parameters as in the caption of Fig.~\ref{fig5}.}
		\label{fig6}
	\end{center}
\end{figure}

Figure~\ref{fig6} provides a further interpretation of the two regimes. As long as $B_{\rm N}$ does not exceed half the distance between the modes, the sweep of $B_{x}$ leads to the plateaulike behavior with discrete precession frequencies at $\omega/\omega_{\text{R}}$ and fast switching between them at $B_{x}=(K+0.5)B_0$~\cite{Zhukov2018,Evers2021}, see Fig.~\ref{fig6}(a). It can be explained by a semiclassical description using the average spin concept and taking into account the variation of the electron-nuclear feedback strength, see below in Sec.~\ref{subsec:feedback}~\cite{Zhukov2018,KoptevaPSSB19}. The feedback is maximal at the plateau centers ($B_{x}=KB_0$), leading to a reduction of the nuclear spin fluctuations~\cite{Evers2021}, while the feedback is reduced between the plateaus. Here, the nuclear polarization becomes unstable with respect to the neighboring precession mode, which is observed as a frequency jump to the next plateau.

Once $B_{\rm N}$ reaches values higher than $B_0/2$, then the frequency plateau (i.e. the locked frequency mode) can extend beyond the $B_0/2$ range. An example is given for a sweep start field of $B_{\rm s}=4.77$\,T so that $\omega=38\,\omega_{\text{R}}$ in Fig.~\ref{fig6}(b). Here, the first integer mode remains over the whole $B_0/2$ range and even beyond. Above the value of $B_0$, the reduced feedback leads to a partial switch of a QD fraction to the next frequency modes at $B_{x}=K B_0$ ($K>38$), as supported by the experimental results presented in Fig.~\ref{fig1}(c).

\subsection{Feedback strength of electron-nuclear spin interactions. Nuclear spin fluctuations.}\label{subsec:feedback}

The amplitude dependence of the modes, which become locked, could be expressed in terms of the transverse electron spin relaxation time $T_2$. This time is defined by the nuclear spin fluctuations and can be written \cite{MerkulovEfrosRosen,Glazov_book} as
\begin{eqnarray}\label{eq:NF5}
T_\text{2} = \frac{\hbar}{\mu_\text{B}g_{\text{e}}\sqrt{\langle \delta B_\text{N}^2 \rangle} } = \frac{\hbar N}{A\sqrt{\langle \delta I_\text{N}^2\rangle}},
\end{eqnarray}
where
\begin{eqnarray}\label{eq:NF6}
\sqrt{\langle\delta I^2_\text{N}\rangle} = \sqrt{\frac{I(I+1)N|\lambda_0|}{|\lambda|}}.
\end{eqnarray}
$\langle \delta I^2_\text{N}\rangle$ is the characteristic square of nuclear spin fluctuations~\cite{Glazov_book}, depending on the stability of the solution of  Eq.~(\ref{eq:DNP}), defined via the stability parameter $\lambda$. As one can see from Eq.~(\ref{eq:NF6}) the smaller the stability parameter, the greater the magnitude of the nuclear spin fluctuations $\sqrt{\langle\delta I^2_\text{N}\rangle}$.

The stability parameter \cite{Korenev2011}
\begin{eqnarray}\label{eq:NF7}
\lambda = -\frac{T_\text{1e} + T_\text{1L}}{T_\text{1e}T_\text{1L}} \left(1 - \frac{Af_\text{N}\bar{Q}}{\hbar} \frac{\delta \langle S_x \rangle}{\delta \omega_\text{N}}\right),
\end{eqnarray}
where $|\lambda_0| = (T_\text{1e,0} + T_\text{1L})/T_\text{1e,0}T_\text{1L}$ is the stability parameter in case of zero nuclear polarization [$T_\text{1e,0} = T_\text{1e}(\omega_\text{N} = 0)$].

$\lambda$ depends on the nuclear spin polarization time Eq.~(\ref{eq:T1N}), which depends on $B_\text{N}$. As given in the transcendental Eq.~(\ref{eq:DNP}), the feedback strength is defined via the stability parameter. Thus, the larger the Overhauser field, the smaller the feedback strength, and the less stable the solution to Eq.~(\ref{eq:DNP}) which results in increased nuclear spin fluctuations and a decreased transverse electron spin relaxation time according to the results of Ref.~\cite{MerkulovEfrosRosen}. The model, therefore, elucidates that the Overhauser field which allows for the \textit{mode dragging} in each QD is also responsible for the \textit{mode pickup} in the QD ensemble by assigning an individual $T_\text{1e}$ to individual QDs with an optical detuning $\Delta$.

\begin{figure}[t]
	\begin{center}
		\includegraphics[trim=0mm 1mm 0mm 1mm, clip, width=1.05\columnwidth]{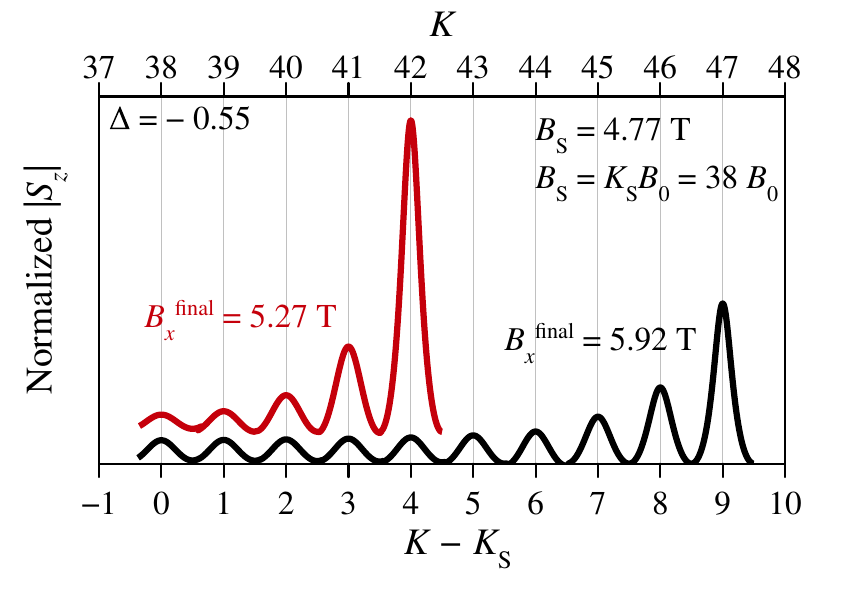}
		\caption{Electron spin polarization spectra ($S_z$) calculated at $B_{x}^\text{final} = 5.27$\,T (red curve) and $B_{x}^\text{final} = 5.92$\,T (black curve) with the field sweep starting at $B_{\rm s} = 4.77$\,T. The $x$ axis is labeled by the effective mode number $K$ starting from the mode number $K_{\rm{s}}$ at the starting field $B_{\rm{s}}$. The normalized curves are vertically shifted for clarity. Used parameters are in the caption of Fig.~\ref{fig5}.}
		\label{fig7}
	\end{center}
\end{figure}

As an example, Fig.~\ref{fig7} demonstrates the simulation of the mode distribution through the corresponding spin polarization for two different sweep ranges with the same starting field ($B_{\text{s}}=4.77\,\text{T}=38\, B_0$). Here we take into account the varying feedback strength, as one can understand through the mode-width shown in the theoretical spin polarization spectra, Fig.~\ref{fig7}. The red curve shows the spin polarization spectrum for a $B_{x}$ sweep from $B_{\rm s}$ to $B_{x}^\text{final} = 5.27$\,T. The last mode number 5 requires zero nuclear spin polarization and has the narrowest width (leading to the longest spin relaxation time $T_2$) as well as highest spin polarization. The lower modes have a reduced amplitude and a larger width. This is additionally demonstrated by performing a longer $B_{x}$ sweep (the black spectrum). It is important to note that the total integral of the spin polarization is the same in both spectra. The decay of the amplitudes with increasing nuclear polarization (lower numbers on the $x$-axis in Fig.~\ref{fig7}) is caused by the decrease of the feedback and the corresponding increase of the nuclear fluctuations. The results observed in experiment do not allow us to determine the spin relaxation time for each mode directly. However, the term $S_0$ in Eq.~(\ref{eq:S}) includes the spin coherence time and shows that the amplitude distribution follows the calculations quite well, see Fig.~\ref{fig1}(b) and Fig.~\ref{fig7}.

\begin{figure*}[t]
\begin{center}
		\includegraphics[width=2\columnwidth, trim=5mm 7mm 5mm 5mm]{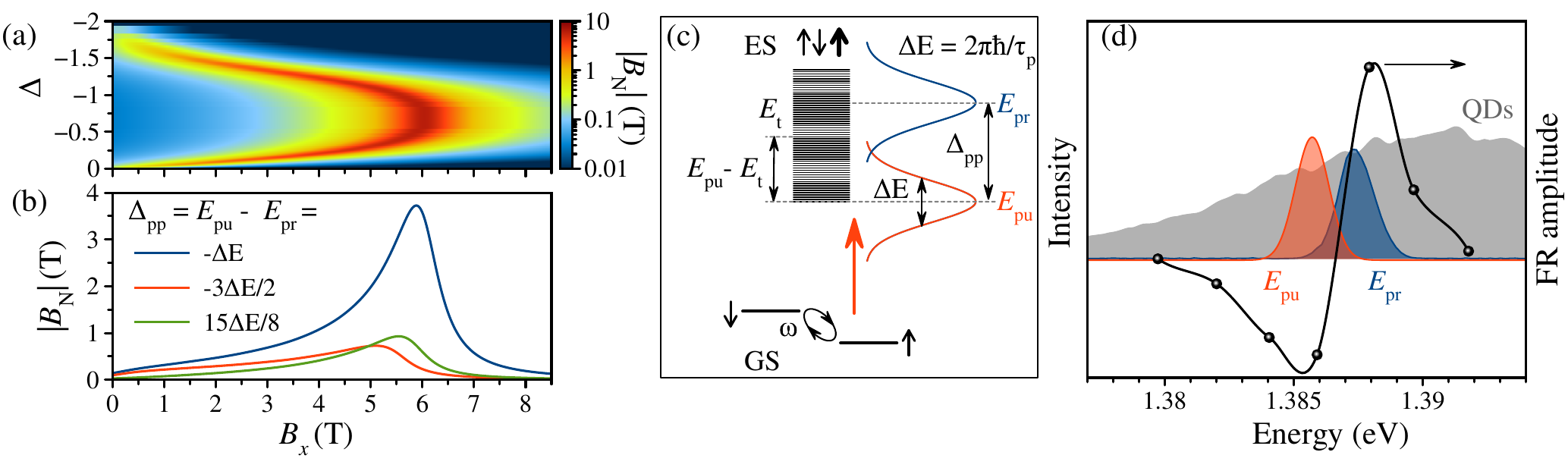}
		\caption{(a) Contour plot of $B_{\rm N}$ vs $\Delta$ and $B_{x}$ in logarithmic scale. It repeats Fig.~\ref{fig5}(c). (b) Dependence of the Overhauser field $B_{\rm N}$ on $B_{x}$ for an optically anharmonic ensemble. The parameter $\Delta_\text{pp}$ corresponds to the difference in optical energies of pump and probe. (c) Scheme of optical trion transitions (black lines) excited by a pump with spectral width $\Delta E$ (red line). The probe pulse (blue line) has the optical detuning $\Delta_\text{pp}$ from the pump pulse. The probe tests the trion resonances in the spectral overlap region of the pump and probe. The resulting precession frequency of the electron spin in the ground state (GS) is shown at the bottom. (d) The gray shaded area shows the QDs PL (possible trion energies), the blue shaded pulse gives the fixed probe spectrum, and the red one depicts the pump spectrum mainly used in the experiment. The black dots (the line is a guide to the eye) show the FR amplitude for different central pump energies. The FR amplitude follows a dispersive dependency on the probe energy.}
		\label{fig8}
	\end{center}
\end{figure*}

The model also allows us to give a figurative explanation of the {\it mode pickup}. Each QD can generate a maximal Overhauser field which compensates the change of the external field, depending on the optical detuning [see Fig.~\ref{fig5}(b)]. Once this individual maximal Overhauser field is reached, it breaks down and the resident electron spin precesses with the frequency which fits the external field without any additional Overhauser field~\cite{Korenev2011,Krebs2010}. The Overhauser field will build up once again with additional changes of the external field. As the ensemble consists of QDs with inhomogeneous trion energy distribution, the breakdown field is inhomogeneous and thus leads to the simultaneously present modes in the studied ensemble. In the next subsection we discuss the effect of inhomogeneity and finite spectral pulse widths in more details.

\subsection{Measurement of Faraday rotation signal}
A pump pulse with spectral width $\Delta E = 2\pi\hbar/\tau_p$ covers a set of trion resonances ($E_{\rm t}$) optically detuned from the pump optical energy. Each optically-detuned trion resonance results in the Overhauser field shown in color in Fig.~\ref{fig8}(a). Due to the dependence of the leakage factor on the external magnetic field, the Overhauser field also depends on the external magnetic field ($B_{x}$). We consider that the probe pulse with spectral width ($\Delta E$) tests a set of optical resonances excited by the pump pulse, see Fig.~\ref{fig8}(c). If the optical detuning between pump and probe ($\Delta_\text{pp} = E_{\text{pu}}-E_{\text{pr}}$) is between zero and $\Delta E$, then all negatively detuned trion resonances participate in the resulting pump-probe signal. Therefore, the average of the Overhauser field distribution over the negative optical detunings as a function of the external magnetic field takes the form of the blue curve in Fig.~\ref{fig8}(b). In this spectral range the majority of QDs fulfills $B_\text{N} > B_0/2$, so that mode dragging and pickup can be observed in the experiment. If the probe is detuned away from the pump to the value $\Delta_\text{pp} = -3\Delta E/2$, only detunings in the range $-2<\Delta < -1$ contribute to the signal and the Overhauser field distribution becomes smaller in amplitude. Additionally, if the probe is positively optically detuned from the pump energy ($\Delta_\text{pp} = 15\Delta E/8$), only the trion resonances with optical detuning in the range $-0.5<\Delta < 0$ contribute to the signal and the Overhauser field also has a small amplitude. For the last two cases, the biggest part of QDs fulfills $B_\text{N} < B_0/2$, and the dependence of the electron precession frequency on the external magnetic field would only demonstrate a steplike behavior.

One should note, that the Overhauser field for positive optical detunings ($\Delta_\text{pp}>0$) is smaller than for negative optical detunings~\cite{KoptevaPSSB19}. A positive optical detuning leads to a much weaker feedback strength and strong nuclear spin fluctuations. Additionally, the positive optical detunings show a frequency locking to the values at $K - 1/2$. Therefore, we expect from the model that QDs with positive detunings do not show dragging effects as experimentally shown~\cite{Hoegele2012}.

$\Delta_\text{pp}$ additionally affects the amplitude of the Faraday rotation signal as shown theoretically in Fig.\,6 of Ref.~\cite{Yugova2009}. The spectral dependence of the Faraday rotation signal amplitude on $\Delta_\text{pp}$ for our experiment is shown in Fig.~\ref{fig8}(d). It is in good agreement with the theoretical result of Ref.~\cite{Yugova2009}. However, the signal amplitude dependence is not covered by the model here and does not affect the nuclear polarization discussed in this work.

\section{Conclusions} 
We have observed experimentally two effects of electron-nuclear spin interaction in a self-assembled, singly charged (In,Ga)As/GaAs QD ensemble under $1\,$GHz pulsed excitation: \textit{mode pickup} and \textit{mode dragging}. The first effect expresses the collection of all commensurate modes whose resonance is met during the transverse external magnetic field sweep. Upon reaching a commensurate frequency $\omega_{\text{R}}$, the corresponding mode is added to the precession spectrum. The second effect expresses the reduced average precession frequency of the QD ensemble relative to the nominal value expected from the applied $B_{x}$. E.g., the initially contributing mode corresponding to the start field is shielded from the varied external field, and stays dominant during the whole field sweep, even though its amplitude slowly drops.

The achieved Overhauser field $B_{\rm N}=1.5$\,T is not the limit for this method. The next step will be to provide benchmark measurements for identifying the limiting factors. Taking into account the inherent strain environment of our QDs, the noncollinear type of interaction mediated by the quadrupolar moments of the nuclei is expected to play an important role~\cite{Urbaszek2013}. It is expected to drive the nuclear spin diffusion under pulsed-laser excitation, which has an effect on the maximally achievable nuclear spin polarization~\cite{Ladd2010a}. These considerations are out of the scope of this paper and will be presented elsewhere. The contour plot in Fig.~\ref{fig5}(c) gives a hint on the best way to achieve the highest nuclear polarization: following the maximal nuclear polarization rate (or highest $B_{\rm N}$ for each detuning) by simultaneous adjusting of the optical detuning $\Delta$ and the magnetic field $B_{x}$.

\begin{acknowledgments}
We acknowledge financial support by the Deutsche Forschungsgemeinschaft in the frame of the International Collaborative Research Center TRR 160 (Project A1) and the Russian Foundation for Basic Research (Grant No. 19-52-12059). A.G. acknowledges support by the BMBF-project Q.Link.X (Contract No.16KIS0857). We acknowledge the supply of the quantum dot sample by D. Reuter and A.~D. Wieck.
\end{acknowledgments}

\appendix
\section{Long term dynamics of the electron-nuclear spin system}\label{app:times}

\begin{figure}[t]
\begin{center}
		\includegraphics[width=\columnwidth, trim=2mm 2mm 2mm 2mm]{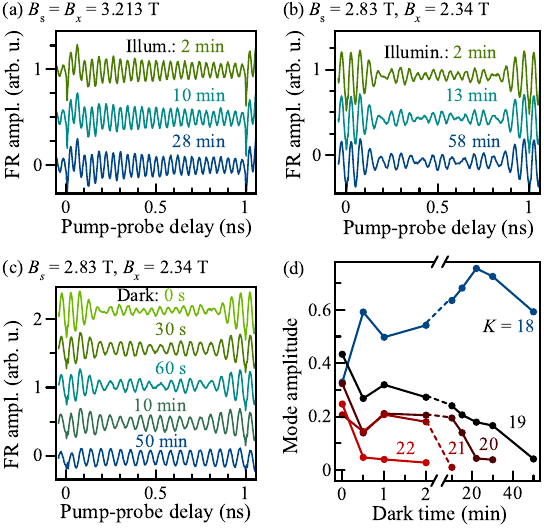}
		\caption{Long-term behavior of the electron-nuclear spin system subject to pulses with 1\,GHz repetition frequency. (a) Pump-probe spectra at constant $B_{x}$ for uninterrupted illumination during the indicated time. No additional modes are introduced by long illumination alone. (b) Pump-probe spectra with uninterrupted illumination. The system is prepared by sweeping from the start field $B_{\rm s}=2.83\,$T to the measurement field $B_{x}=2.34\,$T under illumination. The beating pattern remains stable for uninterrupted illumination. (c) Pump-probe spectra of the system prepared by sweeping the external field under illumination from $B_{\rm s}=2.83\,$T to $B_{x}=2.34\,$T. After each sweep the illumination is blocked for the indicated dark time. The number of nodes in the beating patterns is reduced from 4 ($0\,$s, without dark time) to 0 ($50\,$min dark time). (d) The amplitude of contributing modes is plotted against the dark time. With increasing time, modes requiring a high nuclear polarization fade faster, the higher the associated polarization.}
		\label{fig9}
	\end{center}
\end{figure}

We study the long-term behavior and stability of the electron-nuclear spin system in Fig.~\ref{fig9}. In a first step, we rule out that continuous illumination at a constant magnetic field leads to the emergence of additional precession modes via dynamic nuclear polarization, see Fig.~\ref{fig9}(a). Here, the external field is swept to $B_{\rm s}=B_{x}=3.213\,$T in darkness and the laser is unblocked. The single mode precession remains unchanged throughout $28\,$min of uninterrupted pulsed illumination. In a second step, we examine the stability of the beating pattern for continuous illumination. Figure~\ref{fig9}(b) shows the pump-probe trace directly after preparation by a field sweep from $B_{\rm s}=2.83\,$T to $B_{x}=2.34\,$T (dark green). The beating pattern persists up to one hour of uninterrupted pulsed illumination.

In a next step, we monitor the decay of the precession modes in darkness. The external field is decreased from $B_{\rm s}=2.83\,$T to $B_{x}=2.34\,$ with continuous illumination by the $1\,$GHz laser as before (light green). We observe four beating nodes in the beating pattern (the modes corresponding to $K=22$ to 18) in the time-resolved Faraday rotation signal as shown in Fig.~\ref{fig9}(c). To observe the decay of these modes we sweep the external field down to $0\,$T without illumination, illuminate the sample without external field for $20\,$s, block the illumination again and sweep the field to $B_{\rm s}=2.83\,$T. Using continuous pulsed illumination, we sweep the field down to $B_{x}=2.34\,$T and block the laser for $30\,$s. Subsequently the next pump-probe trace is acquired (dark green). The process is repeated for increasing times in darkness and the resulting Faraday rotation traces are analyzed as in the main text. The amplitudes of the modes are plotted against the time in darkness in Fig.~\ref{fig9}(d). The modes to which a nuclear magnetic field contributes decay faster the larger this nuclear field is: The mode which only requires a minimal additional field is contributing more strongly over time. We therefore conclude that with increasing dark time the developed nuclear field decays within seconds to minutes which is known to be the typical timescale for the depolarization of nuclear spins in (In,Ga)As/GaAs QDs~\cite{Glazov_book}.

\section{Maximal Overhauser field}\label{app1}

To estimate the maximal possible Overhauser field in the ensemble of (In,Ga)As/GaAs QDs we use the expression~\cite{MerkulovEfrosRosen}:
\begin{align}
B_{\text{N,max}} = \sum\limits_{\text{j}} \frac{I_{\text{j}}A_{\text{j}}\chi_{\text{j}}n_{\text{QD,j}}}{g_{\rm e} \mu_{\rm B}} = 6.23\,\text{T}\label{eq:max-field}
\end{align}
with the index $j$ running over all isotopes in the QDs, the nuclear spin $I_{\text{j}}$, the hyperfine constant $A_{\text{j}}$, the natural abundance $\chi_{\text{j}}$, and the respective fraction of the nuclear species in the QD composition $n_{\text{QD,j}}$. The electron $g$ factor is $g_\text{e}=-0.56$. The numerical values are given in the Table~\ref{tab:nuclear-spin}.
Note that especially the fractions of In ($35\,\%$) and Ga ($65\,\%$) represent rough estimates based on the findings in Ref.~[\onlinecite{Petrov2008}]. The maximal field $B_{\text{N,max}}$ could range between $4.3 \,$T for $0 \,$\% In up to $9.9\,$T for $100\,$\% In in the QDs.

\begin{table}[h]
	\centering
	\caption{Parameters of the nuclear spin system in the studied QD sample. The nuclear spin $I_{\text{j}}$, the hyperfine interaction $A_{\text{j}}$, the natural abundance $\chi_{\text{j}}$, and the fraction of the nuclear species $n_{\text{QD,j}}$ in the QD composition are given.}
	\begin{tabular}{ccccc}
		& $I_{\text{j}}$ \cite{Haynes2012} & $A_{\text{j}}, \upmu$eV \cite{Chekhovich2017} & $\chi_{\text{j}}, \%$ \cite{Haynes2012} & $n_{\text{QD,j}}$,\%\\
		\hline
		$^{113}\text{In}$ & 4.5 & 56 & 4 & \multirow{2}{*}{35} \\
		$^{115}\text{In}$ & 4.5 & 56 & 96 &\\
		$^{69}\text{Ga}$ & 1.5 & 43 & 60 & \multirow{2}{*}{65} \\
		$^{71}\text{Ga}$ & 1.5 & 55 & 40 &\\
		$^{75}\text{As}$ & 1.5 & 44 & 100 & 100 \\
	\end{tabular}
	\label{tab:nuclear-spin}
\end{table}
%----------------------------------
%merlin.mbs apsrev4-1.bst 2010-07-25 4.21a (PWD, AO, DPC) hacked
%Control: key (0)
%Control: author (0) dotless jnrlst
%Control: editor formatted (1) identically to author
%Control: production of article title (0) allowed
%Control: page (1) range
%Control: year (0) verbatim
%Control: production of eprint (0) enabled
%
%----------------------------------

\end{document}